\documentclass[aps,pre,reprint,floatfix]{revtex4-1}
\pdfoutput=1
\usepackage{graphicx}
\usepackage{amssymb,amsfonts,amsmath}
\usepackage{psfrag}

\newcommand{\xhat}{\hat{x}}
\newcommand{\yhat}{\hat{y}}
\newcommand{\nhat}{\hat{n}}

\newcommand{\nut}{\nu^S_\text{T}}
\newcommand{\Pt}{\mathbf{P}_\text{T}}
\newcommand{\pt}{{P_\text{T}}}
\newcommand{\nm}{n_\text{m}}
\newcommand{\nss}{n_\text{ss}}
\newcommand{\dl}{\mathbf{d}_\text{L}}
\newcommand{\dr}{\mathbf{d}_\text{R}}

\begin{document}
\title{Topological modes bound to dislocations in mechanical metamaterials}
\author{Jayson Paulose}
\affiliation{Instituut-Lorentz, Universiteit Leiden, 2300 RA Leiden, The Netherlands}

\author{Bryan Gin-ge Chen}
\affiliation{Instituut-Lorentz, Universiteit Leiden, 2300 RA Leiden, The Netherlands}

\author{Vincenzo Vitelli}
\email{vitelli@lorentz.leidenuniv.nl}
\affiliation{Instituut-Lorentz, Universiteit Leiden, 2300 RA Leiden, The Netherlands}

\maketitle

{\bfseries
 Mechanical metamaterials are artificial structures with unusual 
 properties, such as negative Poisson ratio, bistability or tunable 
 vibrational properties, that originate in the geometry of their 
 unit cell~\cite 
 {95181,Schenk26022013,PhysRevLett.110.215501,Sun31072012,Shan2014}.
 At the heart of such unusual behaviour is often a soft 
 mode: a motion that does not significantly stretch or compress the 
 links between constituent elements. When activated by motors or 
 external fields, soft modes become the building blocks of robots
and smart materials. Here, we demonstrate the existence of 
 topological soft modes that can be positioned at 
 desired locations in a metamaterial while being robust against a 
 wide range of structural deformations or changes in material
 parameters~\cite{Kane2013,Prodan2009,Vitelli31072012,Chen2014,Vitelli2014}. These 
 protected modes, localized at dislocations in deformed kagome and square lattices, are the mechanical 
 analogue of topological states bound to defects in electronic 
 systems~\cite{Stern2008,Ran2009,Teo2010,Juricic2012}. We create physical 
 realizations of the topological modes in prototypes of kagome lattices 
 built out of rigid triangular plates. We show mathematically that 
 they originate from the interplay between two 
 Berry phases: the Burgers vector of the dislocation and the 
 topological polarization of the lattice. Our work 
 paves the way towards engineering topologically protected 
 nano-mechanical structures for molecular robotics or information 
 storage and read-out.
}

\begin{figure}
    \includegraphics{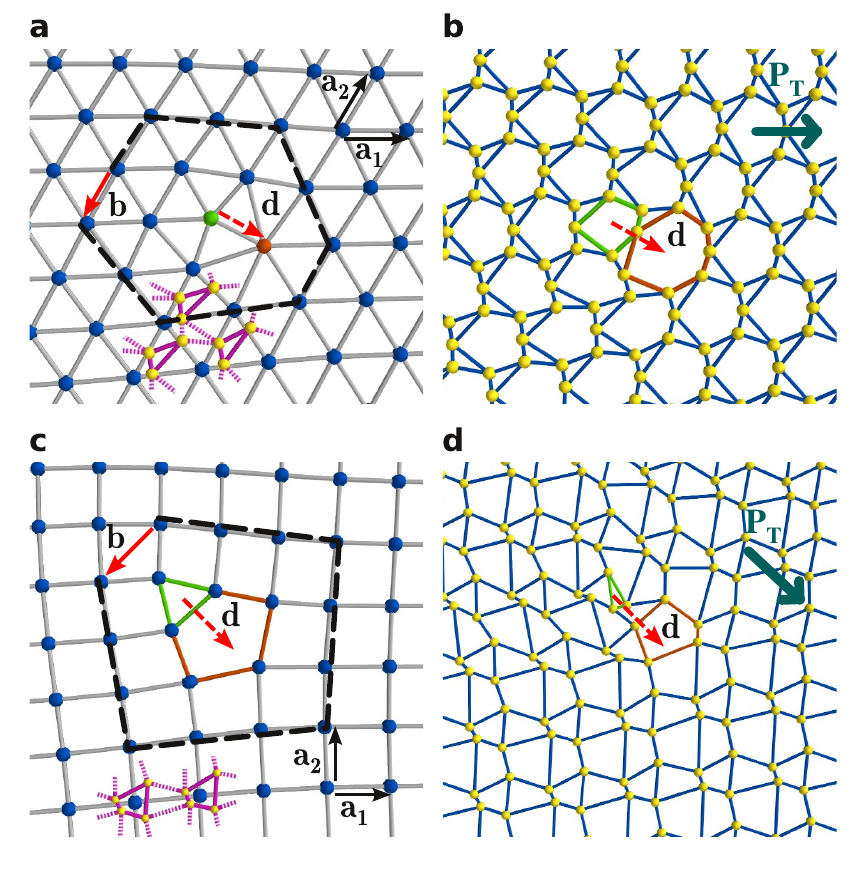}
    \caption{{\sffamily 
    {\bfseries Dislocations in polarized isostatic lattices.}
    {\bfseries a,} Hexagonal lattice with primitive vectors 
    $\{\mathbf{a}_1,\mathbf{a}_2\}$.
    The lattice includes an elementary dislocation, 
    consisting of a five-coordinated point (green) 
    connected to a seven-coordinated point (orange) in 
    an otherwise six-coordinated lattice (blue points). The 
    Burgers vector $\mathbf{b}=-\mathbf{a}_2$ is the deficit in 
    a circuit (black dashed line) that would have been closed in a 
    defect-free lattice. Rotating this vector by $\pi/2$ gives the 
    corresponding dipole moment vector $\mathbf{d}$, which 
    connects the five-coordinated point to the seven-coordinated point.  
    Decorating each unit cell with a three-atom basis (yellow  
    points and magenta bonds) produces a dislocated deformed kagome lattice which 
    contains only four-coordinated points. Three copies of the three-atom 
    basis are shown; solid bonds connect points within the 
    same unit cell whereas dashed bonds connect points belonging to different 
    cells.
    {\bfseries b}, Deformed kagome lattice obtained from decorating the 
    triangular lattice in {\bfseries a}, thus incorporating a dislocation with 
    the same dipole moment $\mathbf{d}$. The five- and seven-coordinated 
    points in the underlying triangular lattice translate into plaquettes 
    bordered by five (green) and seven (orange) bonds respectively, whereas 
    all other points in the triangular lattice translate to plaquettes 
    bordered by six bonds (blue) in the decorated lattice. The 
    topological polarization $\Pt=\mathbf{a}_1$, calculated in 
    Ref.~\onlinecite{Kane2013}, is also shown.
    {\bfseries c}, Square lattice with primitive vectors 
    $\{\mathbf{a}_1,\mathbf{a}_2\}$ (black arrows) which incorporates a 
    dislocation, consisting of a three-coordinated plaquette 
    (bordered by green bonds) adjacent to a five-coordinated plaquette 
    (bordered by orange and green bonds), with Burgers vector $\mathbf{b} = 
    -(\mathbf{a}_1+\mathbf{a}_2)$. The associated dipole moment $\mathbf{d}$ connects the three- 
    and five-coordinated plaquettes. Decorating each point with the four-point 
    unit cell (yellow points and magenta bonds) gives the distorted square 
    lattice in {\bfseries{d}} which incorporates a dislocation of the same 
    dipole moment, and has a nonzero topological polarization 
    $\Pt = \mathbf{a}_1-\mathbf{a}_2$.
    }} \label{fig_decoration}
\end{figure}

Central to our approach is a simple insight: mechanical structures 
on length scales ranging from the molecular to the architectural can 
often be viewed as networks of nodes connected by links~\cite{Maxwell}.
Whether the linking components are chemical bonds or 
metal beams, mechanical stability depends crucially on the number of 
constraints relative to the degrees of freedom. When the degrees of 
freedom exceed the constraints, the structure displays excess zero 
(potential) energy modes. Conversely, when the constraints exceed 
the degrees of freedom, there are excess states of 
self-stress---balanced combinations of tensions and compressions of the 
links with no resultant force on the nodes. The generalized 
Maxwell relation~\cite{Calladine1978} stipulates that the index $\nu$ given by 
the difference between the number of zero modes, $\nm$, and the number of  
states of self-stress, $\nss$, is equal to the number of degrees of 
freedom $N_\text{df}$ minus the number of constraints $N_\text{c}$ 
\begin{equation}
\nu \equiv \nm-\nss = N_\text{df} - N_\text{c}.
\label{eq1}
\end{equation} 

A trivial way to position a zero-energy mode in the interior of a 
generic rigid lattice is to remove some bonds, locally reducing the
number of constraints. Consider, instead, a network that satisfies everywhere the local isostatic condition $N_\text{df} = 
N_\text{c}$ (which precludes bond removal). In this case, zero modes can only be present in conjunction with 
an equal number of states of self-stress, invisible partners from the 
perspective of motion. Isostaticity by itself, however, does not dictate how 
the modes are distributed spatially. Kane and Lubensky~\cite{Kane2013} recently 
introduced a  
special class of isostatic lattices that possesses an additional feature 
called topological polarization.
Much as an electrically polarized material can localize positive 
and negative charges at opposite boundaries, a 
topologically polarized lattice can harbour zero modes or states of 
self-stress at sample edges (whose outward normal is respectively aligned or anti-aligned 
with the polarization).
The edge mode distribution is biased even though both
boundaries are indistinguishable on the basis of local constraint counting. Furthermore,
this bias is insensitive to local variations in bond lengths, spring constants, or node
masses, provided no bonds are cut and the lattice remains rigid in the bulk~\cite{Kane2013}. 

In this Letter, we harness the topological polarization to 
place zero modes in the interior of an isostatic lattice where 
topological defects called \emph{dislocations} are positioned. 
Dislocations are termination points of incomplete lattice rows that have an 
edge-like character. They are characterized by a topological charge called the 
Burgers vector, $\mathbf{b}$, that measures the deficit in {\it any} circuit surrounding 
the dislocation, see Figs.~1a and 1c. A dislocation is composed of a dipole of 
under-coordinated (green) and over-coordinated (orange) points (Fig.~1a) or 
plaquettes (Fig.~1b--d),
whose orientation is obtained upon rotating $\mathbf{b}$ by $\pi/2$. The dipole moment 
$\mathbf{d}$, a vector connecting the under-coordinated point/plaquette to the 
over-coordinated one, points outward from the added strip of material that terminates at the 
dislocation and has a magnitude equal to the strip width. Therefore, $\mathbf{d}$ 
quantifies the orientation and size of the effective ``edge'' created by the dislocation. 

To localize topological modes at these effective edges, we 
need to incorporate the dislocations into polarized lattices without
modifying the local constraint count (i.e. trivial zero modes must be excluded). We demonstrate this construction for
two polarized lattices: a deformed kagome lattice 
introduced in Ref.~\onlinecite{Kane2013} 
and a deformed square lattice. As shown in Fig.~1, both lattices are obtained by decorating a 
2D crystal lattice  
(regular hexagonal and square, with primitive vectors 
$\{\mathbf{a}_1,\mathbf{a}_2\}$ indicated in Figs.~1a and 1c respectively) 
with a multi-atom basis at each unit cell. In the absence of defects, the 
unit cell determines the topological polarization $\Pt$ of the bulk lattice. 
In Appendix A we show that $\Pt = \mathbf{a}_1-\mathbf{a}_2$ for the deformed square lattice while it was shown in Ref.~\onlinecite{Kane2013} that 
$\Pt = \mathbf{a}_1$ for the deformed kagome lattice.
Dislocations in the undecorated lattice carry over to the polarized lattice 
and, when appropriately chosen, produce a lattice that is four-coordinated 
everywhere (Figs.~1b and 1d).
See Appendix B for more details of the construction.

\begin{figure*}
    \includegraphics[width=0.8\textwidth]{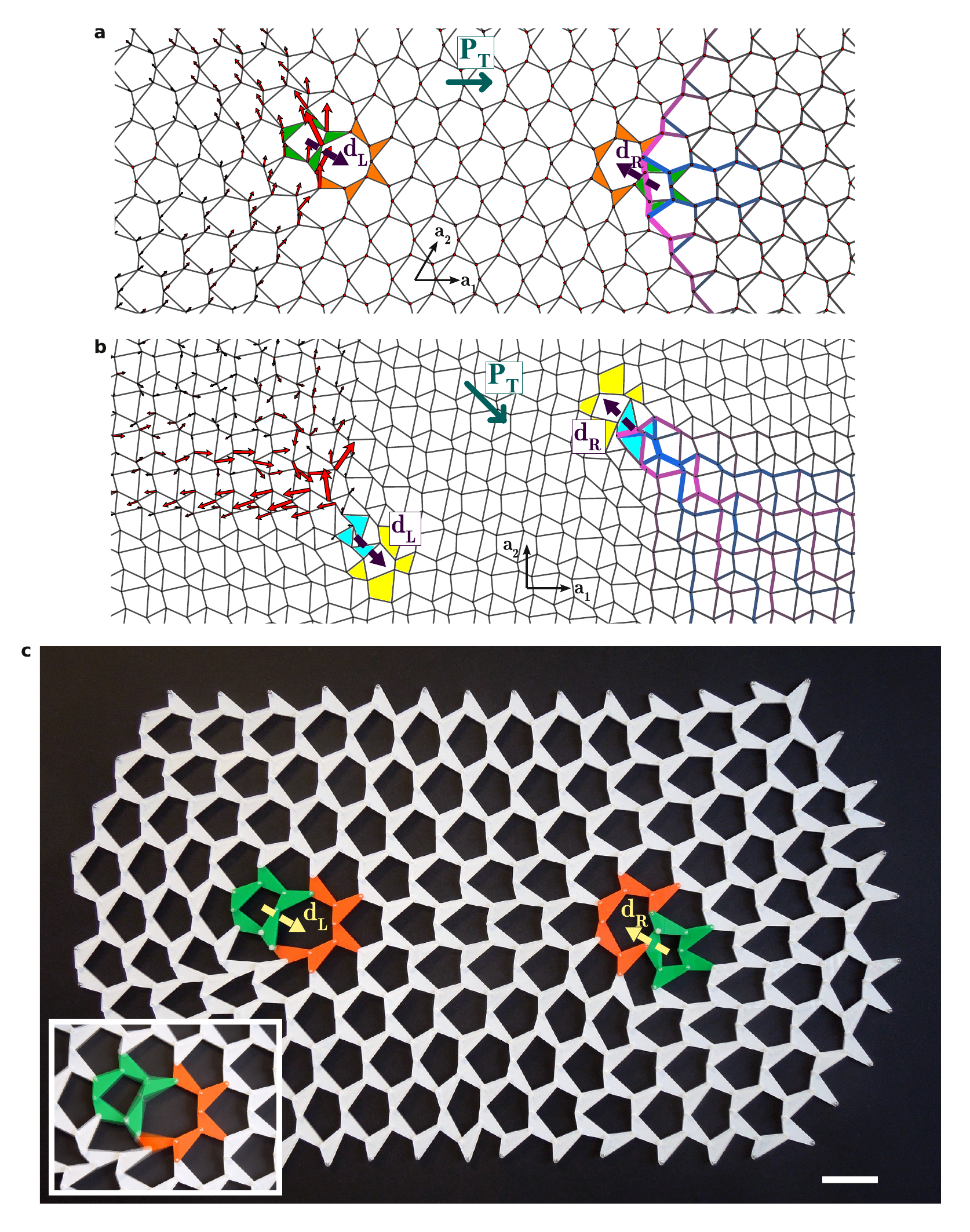}
    \caption{{\sffamily 
    {\bfseries Mechanical modes localized at defects.}
    {\bfseries a,} Visualization in a deformed kagome lattice of a numerically-obtained 
    low-energy soft mode (red arrows, showing direction and relative 
    amplitude of allowed displacements) and an approximate state of 
    self-stress (thickened bonds, showing bond forces in magenta (+) and 
    blue (-) that cancel each other) associated with a pair of 
    dislocations with equal and opposite dipole moments $\dl$ and $\dr$.
     The dislocations are in the interior of a lattice with periodic 
     boundary conditions that is perfectly isostatic. Only a small region of the lattice is shown.
     Each dislocation consists of a 
     five-coordinated plaquette (enclosed by green triangles) adjacent to a 
     seven-coordinated plaquette (enclosed by green and orange triangles).
    {\bfseries b,} Section of a deformed square lattice of a numerically-obtained 
    low-energy soft mode and a state of self-stress 
    associated with a pair of 
    dislocations with equal and opposite dipole moments $\dl$ and $\dr$. The 
    visualization method is similar to that in {\bfseries a}. The dislocations are in the 
    interior of a lattice with periodic boundary conditions that is 
    perfectly isostatic. Each dislocation consists of a three-coordinated 
    plaquette (enclosed by cyan plaquettes) near a five-coordinated plaquette 
    (enclosed by cyan and yellow plaquettes). All other plaquettes are 
    four-sided.
    {\bfseries c,} Plastic prototype of a deformed kagome network, built as 
    described in the text. The interior contains two dislocations which 
    reproduce the configuration from the computer model show in  {\bfseries a}. 
    Scale bar  5 cm.
    {\bfseries Inset,} 
    Superposition of three configurations that span the
    range of the free motion associated with the left dislocation.
    }} \label{fig_demo}
\end{figure*}

Having constructed polarized lattices with dislocations, we numerically 
compute their vibrational spectrum by treating each bond as a harmonic spring.
Results are shown in Figs.~2a and 2b for deformed kagome and square networks 
respectively. We use periodic boundary 
conditions, which preserves
isostaticity everywhere but requires a net 
Burgers vector of zero. As a result, dislocations appear in pairs with equal 
and opposite dipole moment. In both networks, the dipole moments of the 
dislocation on the left ($\dl = (2/\sqrt{3})(\mathbf{a}_1-\mathbf{a}_2/2)$ for the
kagome network and $\dl = \mathbf{a}_1-\mathbf{a}_2$ for the square network)  and on
the right ($\dr=-\dl$) are respectively aligned with and against the lattice polarization.
The left dislocation has an associated soft 
mode in both cases, labeled by arrows, whose energy decreases with system size. The 
opposite dislocation is associated with 
an approximate state of self-stress, labeled by colored and thickened bonds following 
Ref.~\onlinecite{Kane2013}.
(See Appendix C for computational details.) These observations are 
consistent with the intuitive interpretation of dislocations as edges oriented by their 
dipole moment.

To assess whether these modes can be observed in metamaterials with 
realistic bonds, joints and boundary 
conditions, we have built prototypes of the 
deformed kagome lattice, composed of rigid triangles laser-cut out of 3 mm thick PMMA sheets. The corner of the triangles are connected by plastic bolts that act as hinges. The boundary points are pinned 
to a flat base by screws, but can pivot freely.     
The design ensures that each internal vertex
has as many constraints as degrees of freedom, satisfying the local 
isostatic condition away from the boundary. Fig.~2c shows such a prototype 
mimicking the dislocation configuration of the computer model from Fig.~2a.
Theoretically, the boundary pinning and the use of rigid triangles push the phonon gap 
to infinity, so that only zero modes can be observed. In practice, 
the prototype has some compliance and 
mechanical play at the pivots. Nonetheless, it is rigid in the bulk, 
as can be verified by unsuccessfully  
trying to move the (white) triangles far from the dislocations, see 
Supplementary Movie 1. 

Despite the differences between the two systems, the soft mode observed in the simulated 
harmonic network survives in the real-world prototype as a collective motion of points near the left 
dislocation. The motion is easily activated by pushing the
hinge joints of the triangles that make up the dislocation (inset to figure~2c
 and Supplementary Movie 2). The motion is not a strict zero mode because it 
 interacts with the pinned boundary of the finite system, but the structural 
 compliance is sufficient for the remnant soft motion to be observed.
In contrast, the dislocation on the right
does not admit displacements in its vicinity and remains rigid (Supplementary 
Movie 3), consistent with the simulations.

To quantify the number and type of modes associated with a
dislocation, an electrostatic analogy proves useful. Once the connectivity of a locally  
isostatic lattice is fixed, the index $\nu$ in equation (1) can be viewed as a 
topological charge, invariant under smooth deformations of the lattice. 
Just as Gauss's law yields the net charge enclosed in a region from 
the flux of the \emph{electric} polarization through its boundary, the net 
value of $\nu$ in an arbitrary portion of an isostatic lattice 
is given by the flux of the \emph{topological} polarization through its 
boundary~\cite{Kane2013}. Upon evaluating the flux of $\Pt$ on a 
contour encircling an isolated dislocation, we obtain 
\begin{equation}
    \label{eqn_count_for_b}
    \nu =\frac{1}{V_\text{cell}}\Pt\cdot  \mathbf{d},
\end{equation} 
where $V_\text{cell}$ is the unit cell area. In Appendix D, we 
present a detailed derivation  
of equation~\eqref{eqn_count_for_b} that accounts for the elastic strains 
around the dislocation. Here, we simply comment on its physical interpretation.

The topologically protected modes arise from a delicate interplay between 
a Berry phase associated with cycles in the Brillouin zone, 
embedded in $\Pt$~\cite{Kane2013}, and the Berry phase of a topological 
defect in real space, represented by its Burgers vector (or 
dipole $\mathbf{d}$). A similar interplay 
dictates the existence of localized electronic modes at dislocations in 
conventional topological insulators~\cite{Ran2009,Teo2010}.
Equation~\eqref{eqn_count_for_b} 
gives $\nu=+1\,(-1)$ for the left (right)
dislocation in the deformed kagome lattice, and $\nu=+2\,(-2)$ in the deformed 
square lattice. The sign of $\nu$ distinguishes zero modes ($+$) from states of self-stress ($-$), 
while its magnitude gives their numbers. For instance, we correctly
predict that the square lattice of Fig.~2b admits two soft modes
localized to the left dislocation (as verified in Supplementary Figure~S1).

The soft modes investigated here have unusual localization properties
as shown in Fig.~3. 
The mode amplitude falls off exponentially along most rays originating at the core of the left dislocation, but the decay length depends on the direction 
of the ray relative to the underlying lattice. There are two special 
directions in each lattice (one of which is highlighted by red circles in 
both Fig.~3a and Fig.~3b) along which the localization is weak. In all other directions, the mode decays 
over much shorter length scales of order a few lattice constants (green circles and 
symbols in Fig.~3). In Appendix E, we show 
that the weak localization directions track lines in momentum space along which the 
acoustic modes vary quadratically, rather than linearly, with the momentum. 
The localization of the approximate state of self-stress behaves similarly, 
with weak directions that are the opposite of those for the soft mode.

\begin{figure*}
    \includegraphics{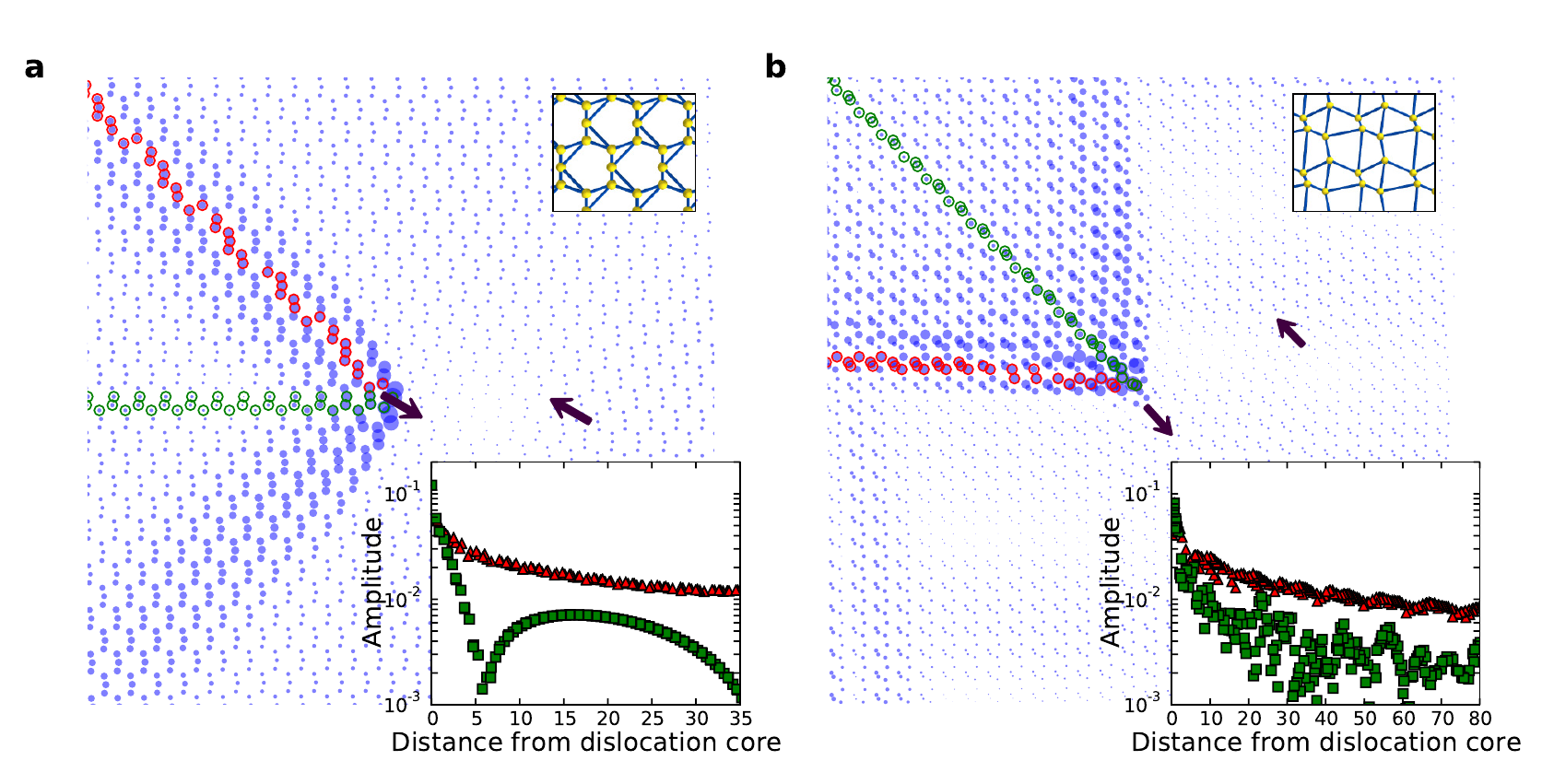}
    \caption{{\sffamily
        {\bfseries Anisotropic localization of the soft mode.}
        {\bfseries a,} Amplitude of the soft mode associated with the left 
        dislocation in the kagome lattice shown in figure~2a,  
        visualized as blue disks whose area is scaled by the displacement magnitude at each 
        lattice point. The dipole moment vectors $d_L$ and $d_R$ (solid 
        arrows) indicate the position and orientation of the dislocations. 
        {\bfseries Inset,} Soft mode amplitude as a function of distance from 
        the left dislocation, along two directions (indicated by red and green 
        circles in the main panel which enclose the lattice points sampled in 
        the red and green curves respectively).
        {\bfseries b,} Same as in {\bfseries a} for the soft mode associated 
        with the left dislocation shown in figure~2b.
    }} \label{fig_localize}
\end{figure*}

The topologically protected modes we have identified could have 
applications across a wide range of systems and length scales. At 
macroscopic scales, isostatic origami structures exist whose deformations are 
restricted to rotations of hinged triangles much as in the kagome lattice~
\cite{tachi2013designing}. At 
the microscale, dislocations could be used for robust information 
storage, with a bit encoded by the presence (+) or absence (-) of a 
topological soft mode, in turn controlled by the orientation of the Burgers 
vector. Such protected bits could be hard-wired into microscopic 
``punch cards'' which can be read out mechanically by probing the region around
dislocations for soft motions (or lack thereof). We also envision molecular robots and smart
metamaterials that exploit the protected modes as activated 
mechanisms. 

\hspace{.1 in}

{\scshape Acknowledgements:} We thank J.~C.~Y.~Teo, R.-J.~Slager, and 
A.~Turner for helpful
discussions, the Leiden University Fine Mechanics Department, 
and the anonymous referees for detailed feedback. This work was funded by FOM 
and by the D-ITP consortium, a program of the Netherlands Organisation for 
Scientific Research (NWO) that is funded by the Dutch Ministry of Education, 
Culture and Science (OCW).   

\hspace{.1 in}

\bibliographystyle{naturemag}
\bibliography{references}

\appendix
\setcounter{figure}{0}
\renewcommand{\thefigure}{S\arabic{figure}}
\section{Calculation of the topological polarization} \label{app_pt}
Here we summmarize the calculation of the topological polarization following 
Ref.~\onlinecite{Kane2013}. The topological polarization
is calculated from the rigidity, or compatibility, matrix $R$ that relates
bond extensions $e_i = R_{ij}u_j$ to site displacements
$u_i$~\cite{graver1993,Calladine1978}.
The first step is to calculate the
Fourier-transformed rigidity matrix $R(\mathbf{k})$ whose
determinant is complex: $\det R(\mathbf{k}) \equiv |\det 
R(\mathbf{k})|e^{i\phi(\mathbf{k})}$ (done with a choice of
unit cell consistent with the calculations in
Ref.~\onlinecite{Kane2013}). Next, the winding numbers,
$n_i=\{n_1,n_2\}$, of $\phi(\mathbf{k})$ are evaluated using
\begin{equation} \label{eqn_windingnumbers}
    n_i=- \frac{1}{2\pi}\oint_{C_i}d\mathbf{k}\cdot\nabla_\mathbf{k}
\phi(\mathbf{k})
\end{equation}
along the two cycles $\{C_1,C_2\}$ of the Brillouin zone
parallel to the reciprocal lattice directions $(\mathbf{b}_1,\mathbf{b}_2)$, 
defined by $\mathbf{a}_i\cdot\mathbf{b}_j=2\pi\delta_{ij}$~\cite{Kane2013}. The
winding numbers are invariants of the gapped lattice; smooth
deformations of the triangle shape do not change $n_i$ unless the 
sites in the lattice lie in a special position and cause 
the phonon gap to close.  We define the topological polarization to be
\begin{equation} \label{eqn_polarization}
    \Pt = \sum_i n_i \mathbf{a}_i.
\end{equation}
This definition of $\Pt$ differs from that of 
Ref.~\onlinecite{Kane2013} by a factor of $V_\text{cell}$ (theirs is strictly
speaking a polarization {\em density}).  Our $\Pt$ is equal to ${\bf
R}_T$ in Ref.~\onlinecite{Kane2013} since the correction ${\bf
r}_0=d\sum_{i\text{ sites}}{\bf p}_i-\sum_{j\text{ bonds}}{\bf p}_j$
(where ${\bf p}_a$ is the position of site / bond $a$) is
equal to zero for our choices of unit cells.
For the deformed square lattice used in the main text, the Fourier-transformed rigidity 
matrix is calculated from the unit cell site positions (given in
section \ref{app_constructsquare}) and bond assignments, 
and the numerically-obtained winding numbers are  $n_1 = 1, n_2 = -1$.
Thus the resulting lattice is characterized by the topological polarization 
$\Pt = \mathbf{a}_1-\mathbf{a}_2$ in terms of the primitive vectors
$\mathbf{a}_i$ indicated in Fig.~1c of the main text.

\section{Constructing isostatic networks with dislocations}

The basic principle for constructing isostatic networks with dislocations 
was outlined in the main text and Fig.~1. Here we provide more
details about the procedure for constructing the lattices used in numerical 
calculations and the real-world prototypes. 

\subsection{Deformed kagome lattice} \label{app_constructkagome}
Following Ref.~\onlinecite{Kane2013}, deformed kagome lattices with nontrivial 
polarization are obtained by decorating a regular hexagonal lattice with a unit 
cell consisting of three points and six bonds. The underlying hexagonal 
lattice is built from the primitive lattice vectors $\mathbf{a}_1 = a\xhat$, 
$\mathbf{a}_2 = (a/2) \xhat + (\sqrt{3}a/2) \yhat$ where $a$ is the lattice 
constant and $\xhat,\yhat$ are unit vectors. Decorating a hexagonal lattice 
with $N$ points and $3N$ bonds produces a deformed kagome lattice with $3N$ 
points and $6N$ bonds. We use the parametrization of 
the unit cell introduced in Ref.~\onlinecite{Kane2013} to describe deformed 
kagome lattices with no asymmetry in the constituent triangles: three numbers 
$(x_1,x_2,x_3)$ quantify the distortion of the  
lines of bonds away from the straight lines in a regular kagome lattice 
($x_i=0$). Any set of three values $x_i$ specifies the positions of the three points in the unit 
cell. Equivalently, each bond in the hexagonal lattice is associated with a 
single point in the kagome lattice. Therefore, for any set of $x_i$, each bond 
in a hexagonal lattice can be replaced with a point whose position is 
calculated from the bond endpoints and the $x_i$ values. Linking up these new 
points following the bond assignments dictated by the kagome lattice unit cell produces the required 
deformed kagome lattice as a decoration of the hexagonal lattice.

Numerical lattices with dislocations are prepared starting from a defect-free hexagonal 
lattice with points connected by harmonic springs of length $a$ and spring 
constant $k$ under periodic boundary 
conditions (the system is a rectangular box with dimensions specified to match 
the size of the finite hexagonal lattice). Using local bond reassignments, a dislocation pair with net zero 
Burgers vector can be prepared at any location; the individual dislocations 
can then be moved around independently of each other through a combination of 
glide and climb steps to obtain the desired separation. The presence of 
defects necessarily introduces strains in the lattice. Between each glide and 
climb step, the network of points and springs is relaxed to their (local) 
minimum energy configuration using a conjugate gradient algorithm. This 
procedure approximates the elastic strains in a
membrane with 2D Young's modulus $2k/\sqrt{3}$ and Poisson ratio 
$1/3$~\cite{Seung1988}. Fig.~1a (main text) shows such an equilibrium 
configuration around a dislocation in a hexagonal lattice.

Once a relaxed hexagonal lattice with dislocations in the required locations 
has been created, the decoration described above is carried out. 
Far away from the defects, the resulting lattice reproduces 
the relative point positions and bond lengths of the deformed kagome lattice
parametrized by the $x_i$. However, in the strained regions of the lattice 
close to the dislocations, the decorated lattice necessarily has triangles 
that are distorted from their ideal strain-free shape prescribed by the $x_i$. 
The dynamical modes of this lattice made out of non-equal triangles can 
still be numerically calculated (following section~\ref{app_numericcalc}) under the
assumption that every bond is tension-free (i.e. by assigning the bond rest 
lengths to be exactly equal to the point separations in the decorated 
lattice), and the resulting spectrum includes approximate zero modes and states of 
self-stress that follow the count of equation~\eqref{eqn_count_for_b}, main 
text. The modes shown in 
Fig.~\ref{fig_si_modeangle} are calculated using this method, which guarantees 
that the unit cell matches the expectation from the $x_i$ far away from the 
defects but at the cost of having irregular triangles near the defect cores.

As mentioned above, elastic distortions of some unit cells are unavoidable 
when preparing kagome networks from hexagonal lattices with defects under 
periodic boundary conditions. However, we can  
reduce the distortions in the triangles making up the kagome network, to bring 
their shapes closer to the ideal shape in a defect-free lattice and make them 
visually similar to the physical prototypes which have of triangles 
of identical shape. We do this \emph{via} the following steps:
\begin{enumerate}
    \item Decorate the numerically relaxed hexagonal lattice as described 
    above, replacing each bond with a point calculated using the choice of 
    $x_i$ and linking neighbours according to the kagome lattice bond 
    assignment. There is no ambiguity in 
    assigning neighbours to the points near the core of the
dislocation, and this assignment
    satisfies the local constraint count of four neighbours to each site and 
    matches the connectivity of the perfect kagome lattice away from the defect.
    \item Assign to each bond in the decorated lattice a rest length that 
    matches the rest length in a defect-free deformed kagome lattice 
    parametrized by $x_i$. This can be unambiguously done for all 
    bonds in the dislocated kagome lattice except for the edges of a single triangle at 
    the core of each dislocation. For this triangle, we make the following 
    choice: we assign bond lengths to its edges in such a way that the ordered 
    bond length assignment in all six-sided plaquettes 
    surrounding the triangle matches that of six-sided 
    plaquettes in the defect-free lattice.
    \item Relax the network of springs to an equilibrium configuration by 
    minimizing the total harmonic spring energy using a conjugate gradient 
    method. Unlike the hexagonal lattice, the kagome lattice has a large space 
    of low-energy deformations that manifest as rotations of the 
    constituent triangles with minimal stretching of the springs, including 
    the soft mode due to the dislocation as well as the long wavelength phonon modes with 
    anomalously low energy described in section~\ref{app_localization}. 
    These modes dominate the relaxation of the network, leading to large 
    sections of the lattice with self-intersecting triangles or configurations 
    that deviate significantly from the uniform kagome lattice even far away 
    from the defect. To reduce the influence of the soft modes on the 
    relaxation, we relax the network under a macroscopic strain imposed by 
    increasing the size of the rectangular box, which has the effect of 
    stiffening the network response. For the system shown in 
    Fig.~2a, a strain of 10\% along the $x$ direction was imposed during the 
    relaxation step. 
    \item The kagome lattice after the spring relaxation step has less shape 
    variation among the constituent triangles and six-sided plaquettes when 
    compared to the lattice before the relaxation step. However, the imposition of 
    a macroscopic strain deforms the unit cell away from the original unit 
    cell defined by the $x_i$. Part of the deformation can be reversed by 
    applying an overall affine deformation to the network that reverses the 
    uniform strain applied in the previous step, returning the periodic box 
    dimensions to their values before the relaxation step. Far away from the 
    dislocations, the network approaches a deformed kagome lattice still built 
    on a hexagonal lattice with primitive vectors $\mathbf{a}_1,\mathbf{a}_2$ 
    unchanged from before, but parametrized by a new set of values 
    $\tilde{x}_i$ that are close to the original $x_i$. 
\end{enumerate}
Upon following the above steps, we obtain a deformed kagome lattice that 
accommodates the strains required by the dislocations without drastic 
differences in the shapes of the constituent triangles. The 
network used for the numerical calculations in Fig.~2a had $(x_1,x_2,x_3) = 
(0.1,-0.1,-0.1)$ before the relaxation step, which resulted in a deformed 
kagome lattice parametrized by $(\tilde{x}_1,\tilde{x}_2,\tilde{x}_3) 
\approx (0.12,-0.06,-0.06)$ away from the dislocation after all steps were 
followed. The changes in the unit cell due to the relaxation are too small to 
affect the topological polarization, which depends only on the sign of the 
$x_i$~\cite{Kane2013}. Therefore, we have a reliable procedure to numerically generate 
deformed kagome lattices, incorporating dislocations under periodic boundary 
conditions, with a desired topological polarization within the class of 
lattices introduced in Ref.~\onlinecite{Kane2013}. As Fig.~2a shows, triangles in the 
network are still distorted by a small amount near the dislocations, because 
the strains around dislocations in a periodic system cannot be fully 
accommodated by triangle rotations alone. To calculate the approximate zero 
modes and states of self-stress, we again redefine the rest lengths of the 
bonds to match the separations between points in the relaxed configuration so that the network is free of 
tensions.

Generating deformed kagome networks with dislocations out of near-identical 
triangles is more straightforward in the physical prototypes. We start with 
identical laser-cut acrylic triangles with circular holes at the corners. The 
center-to-center distances of the holes are equal to the bond lengths in a 
deformed kagome lattice parametrized by $(x_1,x_2,x_3) = (0.1,-0.1,-0.1)$. We 
join pairs of triangles at corners with plastic fasteners, which act
as hinges. Dislocations can be wired in without difficulty, because of the numerous extra degrees of freedom 
introduced by the free boundary of the system. These additional degrees of 
freedom make the structure very floppy, with many unconstrained rotations of 
triangles about their corner pivots available to accommodate the dislocations. 
The floppiness is removed from the prototypes by pinning down the boundary 
points, ultimately producing an overconstrained system. In Fig.~2c, the effect 
of the dislocations on the network is seen in the (bond length-preserving) 
distortions of unit cells. The small size of the finite prototype means that 
the distortions are seen all the way to the boundary, but their amplitude is 
expected to fall off exponentially over a length scale set by the separation 
between the dislocations of equal and opposite Burgers vector. A sample many 
times larger than that constructed could be made to approach a 
distortion-free kagome lattice at the boundary. 

\subsection{Deformed square lattice} \label{app_constructsquare}
Square lattices are another example of isostatic lattices in two dimensions. 
The regular square lattice, although locally isostatic, does not harbour topologically 
protected modes since it does not have a gapped phonon spectrum. The 
presence of straight lines of bonds extending across the lattice gives 
rise to $2 L$ states of self-stress in an $L\times L$ system, and $2 
L$ corresponding zero modes according to equation (1). These show up as 
lines of zero modes along $k_x=0$ and $k_y=0$ in Fourier space. 
However, the gap can be opened up everywhere except at $\mathbf{k}=0$ 
by decorating the square lattice with a suitable $2\times 2$ unit cell that breaks up the 
straight lines of bonds. The topologically nontrivial deformed square lattice 
we use in the main text was obtained by randomly perturbing a regular 
$2\times 2$ unit cell on a regular square lattice with primitive vectors 
$\mathbf{a}_1 = a\xhat$, $\mathbf{a}_2 = a\yhat$. The positions of the four 
points in the unit cell, defined modulo shifts by Bravais lattice vectors, are 
$[(0,0),(0.51a,-0.27a),(0.63a,0.58a),(-0.07a,0.42a)]$. We have checked numerically that the phonon gap for this 
particular unit cell closes only at $\mathbf{k}=0$, which makes $\Pt$ a 
well-defined quantity. See section~\ref{app_pt} for details of the calculation 
of the polarization vector $\Pt$.

The elementary dislocation in a square lattice (with Burgers vector 
equal to $\pm \mathbf{a}_i$) can be viewed as a 
bound pair of two disclinations (defects in lattice orientational order): 
a lattice point with five-fold 
symmetry adjacent to a plaquette with three-fold 
symmetry. The three-fold lattice point disrupts the isostaticity of 
the Bravais lattice, and also of the resulting crystal lattice after 
decoration with the $2\times 2$ basis.
To prepare dislocations that do not change the local balance between degrees of 
freedom and constraints, we instead use pairs of disclinations 
centered on interstitial regions (plaquettes) rather than on lattice points; 
an example is shown in Fig.~1c of the main text. 
Since no lattice point disclinations are used, every lattice point 
still has exactly four bonds emanating from it. Dislocations 
constructed in this manner have one of four possible Burgers vectors 
$\mathbf{b} = \pm  (\mathbf{a}_1 \pm \mathbf{a}_2)$, oriented along 
diagonals of the square plaquettes of the principal lattice. A square 
lattice incorporating such a dislocation can be decorated with a 
$2\times 2$ basis to obtain a lattice that maintains isostaticity at 
each point.

As with the deformed kagome lattices, we numerically create square lattices 
with dislocation pairs by starting from a defect-free regular lattice, creating a 
dislocation pair, and separating the pair using glide and climb steps. Between 
each step we relax the network to its equilibrium structure. To do this, we 
first fill in the next-nearest-neighbour bonds in the structure so that the 
network has non-zero bulk and shear moduli. The result is a regular square 
lattice with elastic strains surrounding the dislocations, as shown in 
Fig.~1c. Upon decorating this lattice with the $2\times 2$ unit cell while 
ignoring the strains, we again obtain distorted unit cells close to the 
defects. We reduce the distortions by following the steps 1--4 outlined for the 
kagome lattice in the previous section, relaxing the network under external 
strain after assigning rest lengths to bonds that match the rest lengths in a 
defect-free lattice with the same $2\times 2$ basis. 
The unit cell does not define the rest lengths of the bonds traversing the 
edge shared by the three-fold and the five-fold 
plaquettes in the base lattice; we arbitrarily reuse rest lengths from the 
defect-free lattice for these bonds. For the square lattice 
reported in the main text, we used a macroscopic strain of 5\% in both 
directions during the relaxation step.

\section{Numerically obtaining and visualizing the soft mode and 
self-stress} \label{app_numericcalc}
The soft mode plotted in Fig.~2a is obtained by constructing and 
numerically diagonalizing the dynamical matrix associated with the 
network. The construction of the network itself follows 
section~\ref{app_constructkagome}. We start with a hexagonal network of 8,317 nodes in a 
periodic box of size $128 a_0 \times 32\sqrt{3} a_0$ which incorporates 
two dislocations with Burgers vectors 
$\pm(\mathbf{a}_1-\mathbf{a}_2)$, separated by roughly seven lattice 
constants. We then decorate and relax the lattice to produce a deformed kagome lattice with 
$(x_1,x_2,x_3) \approx (0.12,-0.06,-0.06)$ in the parametrization of 
Ref.~\onlinecite{Kane2013}, which has a topological polarization 
$\Pt=\mathbf{a}_1$. We obtain a network with 24,951 points and twice as many 
bonds. Figure~2a is a zoomed-in view of a subset of the relaxed network.

The vibrational modes of the structure around this equilibrium are 
eigenvalues of the dynamical matrix $D = R^\dagger R$, where the 
rigidity matrix $R$ relates bond extension $e_i$ (one per bond) with 
node displacement $u_j$ (two per node) via $e_i = R_{ij} u_j$. The 
rigidity matrix is calculated from the point positions in the 
equilibrium configuration (spring constants and node masses are set to 
1.)
The lowest several eigenvalues (squared frequencies) in order of 
increasing magnitude and the corresponding 
eigenvectors of the dynamical matrix
are obtained through sparse matrix diagonalization routines implemented in 
the \texttt{Python} programming language. The two lowest eigenvalue 
modes (with eigenvalues zero up to machine precision) are the trivial 
translations available to the network under periodic boundary 
conditions. 
The first nontrivial eigenvalue (third eigenvalue 
overall in magnitude) is the proposed topological soft mode associated with the 
dislocation with dipole moment $\mathbf{d}_\text{A}$. 

Note that the ``collapse mode'' identified in periodic locally isostatic 
networks by Ref.~\onlinecite{kapko2009collapse} does \emph{not} show up as a zero 
mode in the numerical spectra of the finite networks. This is because the 
collapse mode is associated with a change in the shape of the unit cell, and 
therefore in the shape of the periodic box accommodating a fixed number of 
unit cells. The numerical calculations are carried out assuming a periodic box 
of fixed dimensions. Deforming the finite lattice according to a collapse 
mode would not deform any of the internal bonds of the structure, but it 
would deform boundary-crossing bonds connecting edge nodes with periodic 
images of the primary structure. As a result, the collapse mode has a finite 
energy. If in contrast the periodic box in the simulations were allowed to 
change its shape, the collapse mode of Ref.~\onlinecite{kapko2009collapse} would be 
present in the numerical spectra.

In an infinite 
system, the topological mode would show up as an eigenvector with 
eigenvalue zero; in the finite system, it has a finite frequency due 
to interactions with the other dislocation and with its periodic 
images. However, 
it is well-separated in energy from the succeeding modes. The 
association of the mode with the dislocation on the left is even more apparent in 
the structure of the corresponding eigenvector, which has its highest 
amplitude at nodes in 
a small region of the lattice primarily near the dislocation. In 
contrast, the eigenvectors of the fourth and higher-frequency modes 
have the structure of periodic
acoustic modes that extend over the entire lattice. 

The eigenvector corresponding to each eigenvalue gives 
a displacement (a two-dimensional vector) at each point; 
the magnitude of each displacement vector corresponds to the 
relative amplitude of the mode at that point. (The absolute magnitude 
of the displacements has no physical significance.) In figures~2a--b, we have 
plotted these displacements for the eigenvector corresponding to the 
third eigenvalue as red arrows, each indicating the direction 
and scaled in size by the relative magnitude of the displacement of the 
point at its base. (Red dots replace arrows whose length is less than the dot 
diameter.)

Whereas the eigenvectors of the dynamical matrix correspond to 
normal modes, the eigenvectors of its ``supersymmetric partner'' $\tilde{D}=RR^\dagger$
\cite{Kane2013} correspond to tensions and 
compressions experienced by the individual bonds when the system is
driven with the normal modes. Each component of 
the eigenvector translates to a tension on a particular bond. States of self-stress are 
eigenvectors of the partner matrix with eigenvalue zero; the 
eigenvector corresponding to the smallest eigenvalue of the 
partner matrix is localized (i.e. has appreciable tension values) 
to bonds in the vicinity of the dislocation with dipole moment 
$\mathbf{d}_\text{B}$. This eigenvector is visualized in figure 2b 
by increasing the width of each bond (linearly over the range $w$ to 
$4.6w$, where $w$ is the width of the thinnest bond in the figure) 
by an amount proportional to the magnitude of the tension. The 
colour on each bond is also set by the tension; it interpolates 
between blue (negative tensions) and magenta (positive tensions) with 
gray bonds signifying low or no tension. 
 
The calculation and visualization of modes of the deformed square lattice in 
Fig.~2b follows a similar procedure. We follow the steps outlined in 
section~\ref{app_constructsquare}, starting with a square lattice of 24,766 
nodes in a periodic box of dimensions $192 a_0 \times 128 a_0$, and decorate 
and relax the lattice to get a deformed square lattice of 99,064 points, a 
small region of which is displayed in Fig.~2b with the soft mode and 
approximate state of self-stress visualized as described above. 

\begin{figure*}
    \includegraphics{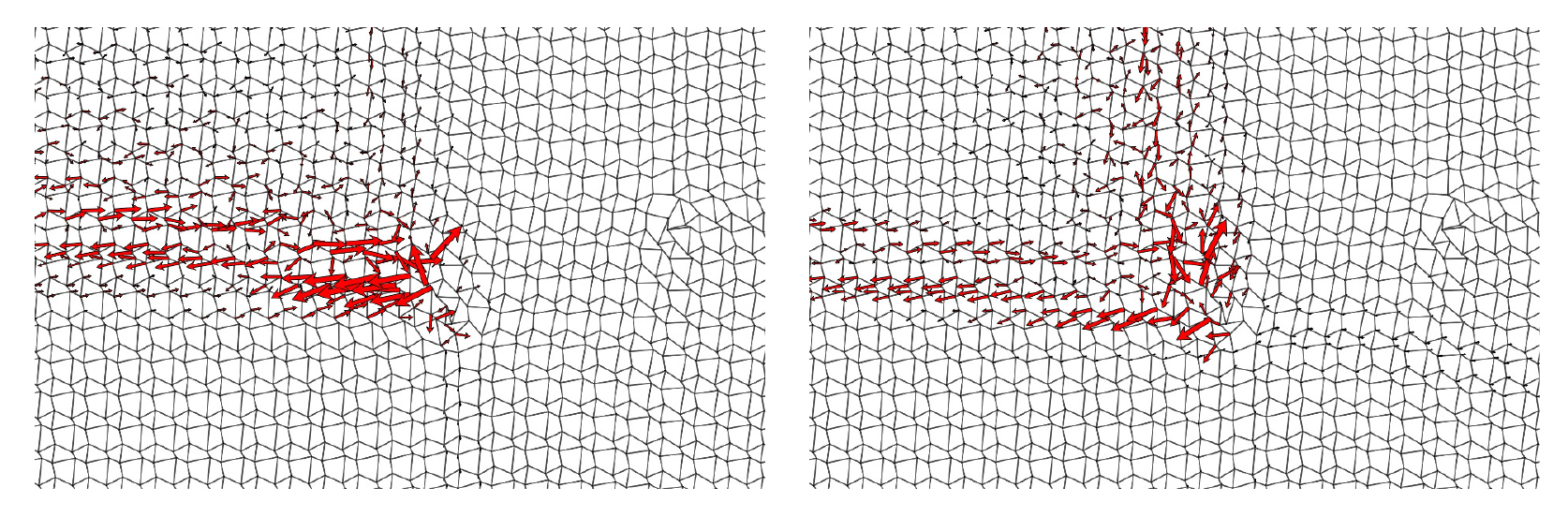}
    \caption{{\sffamily
        Visualization of the eigenvectors of the dynamical matrix of the 
        network in Fig.~2b with the lowest (left) and next-lowest (right) 
        eigenvalues, when the dynamical matrix is evaluated under pinned  
        boundary conditions. Pinning the boundary is implemented by eliminating 
        the degrees of freedom associated with 
        points within $2a$ of the boundary, yielding a rigidity matrix with fewer 
        columns. Such a setup replicates the pinned boundary conditions used 
        in the physical prototypes, and results in a system that is 
        overconstrained, yet exhibits soft modes associated with dislocations 
        of the correct orientation. 
    }} \label{fig_si_twomodes}
\end{figure*}
 
An interesting distinction between the two networks we study 
numerically in the main text is that the mode count associated with the left 
dislocation, as predicted by equation~\eqref{eqn_count_for_b}, is +1 in the 
deformed kagome lattice and +2 in the deformed square lattice. 
Fig.~\ref{fig_si_twomodes} verifies the mode count for the square lattice by 
showing the two lowest eigenmodes of the dynamical matrix for 
the network used in Fig.~2b, but with \emph{pinned} boundary conditions. The 
pinning reduces the interactions between the localized modes and the long-wavelength 
modes of the system and eliminates the wraparound nature of the long ``tail'' 
of the mode in the slow localization direction. This enhances the distinction between the modes 
localized to the left dislocation and the other low-energy modes, which are 
not localized to either dislocation. 
This figure also shows that pinning the boundary, which creates 
an overconstrained system which is no longer strictly isostatic, does not 
eliminate the topological soft modes. This fact is also verified by the physical 
prototype based on the deformed kagome lattice.

\section{Zero mode count of a lattice containing a dislocation} \label{app_count}
Kane and Lubensky~\cite{Kane2013} proved that the number of zero modes
minus the number of states of self-stress
in a lattice subsystem $S$, $\nu^S$, decomposes into a sum
\begin{equation}
\nu^S=\nu_L^S+\nut,
\end{equation}
where $\nu_L^S$ is a ``local'' count and $\nut$ is a ``topological''
count. The local count $\nu_L^S$ is related to the number of unit cells
of the system included in $S$. 
If the unit
cell is chosen so that the term ${\bf r}_0$ is zero, then by choosing the
subsystem $S$ appropriately, $\nu_L^S$ can be arranged to be zero.

If $S$ is bounded by the contour $\cal C$, Kane and
Lubensky~\cite{Kane2013} showed that $\nut$ is given by the formula 
\begin{equation} \label{eqn_topologicalcount}
    \nut = \oint_{\cal C} \frac{d^{d-1}S}{V_\text{cell}}\Pt\cdot\nhat,
\end{equation}
where $\nhat$ is the inward-pointing normal to the boundary, and
$V_\text{cell}$ is the $d$-dimensional volume of the unit cell.  

To compute the zero mode count associated with a dislocation, we need to
evaluate the integral in Eq.~\ref{eqn_topologicalcount} for a path 
surrounding a single dislocation in a lattice with topological polarization
$\Pt$ (as defined in section \ref{app_pt}). We consider paths where
$\nu_L^S=0$ to focus on the effects of the dislocation. 

We compute this integral by taking into a account the deformations in the
lattice due to the dislocation, in the continuum limit.
The dislocation induces a displacement field $\mathbf{u}$ around it. We 
can choose a path sufficiently far away from the dislocation core so 
that the resulting lattice distortions do not change the topological 
winding numbers $n_i$ from their values in the undistorted lattice. 
Then, any variation in $\Pt = \sum_i n_i \mathbf{a}_i$ comes only from 
changes in the local lattice vectors $\mathbf{a}_i$. In the continuum 
limit ($\mathbf{a}$ smaller than the scale of variation of 
$\mathbf{u}$), these are determined by the displacement fields, and as 
a result we have a position-dependent polarization
\begin{equation}
    \pt_i = \pt^0_i+w_{ji}(\mathbf{x})\pt^0_j,
\end{equation}
where 
\begin{equation}
    w_{ji}(\mathbf{x})=\partial u_i/\partial x_j
\end{equation}
is the distortion tensor and $\Pt^0$ is the polarization in the 
undistorted lattice. Meanwhile, the volume of the distorted unit 
cell is
\begin{equation}
    V_\text{cell}(\mathbf{x}) = V_\text{cell}^0(1+w_{ii})
\end{equation}
to linear order in the displacements. Writing out 
Eq.~\ref{eqn_topologicalcount} to linear order in $w_{ij}$, we have
\begin{align}
    \nut &= \oint_{\partial S} \frac{dS}{V_\text{cell}^0} \nhat_i
    \left[\pt^0_i + w_{ji}\pt^0_j - w_{jj}\pt^0_i \right] \\
    &= -\frac{1}{V_\text{cell}^0}\int_S dA\,
    \partial_i\left[\pt_i^0 + w_{ji}\pt^0_j - w_{jj}\pt^0_i \right] \\
    &= \frac{1}{V_\text{cell}^0}\int_S dA\,
    \pt^0_i\left[\partial_i w_{jj} - \partial_j w_{ij}\right], \label{eqn_deriv_1}
\end{align}
where we use the divergence theorem to convert the line integral to an 
area integral (note that $\nhat$ is the \emph{inward} pointing normal
to the contour). However, the distortion tensor is related to the 
Burgers vector via~\cite{ll}
\begin{equation}
    b_k = -\int_S dA\, \varepsilon_{ij}\partial_i w_{jk}.
\end{equation}
Upon applying the antisymmetric tensor to both sides, we get
\begin{align}
    \varepsilon_{km} b_k &= -\int_S dA\, \varepsilon_{ij}\varepsilon_{km}\partial_i w_{jk} \\
    &= \int_S dA\, (\delta_{im}\delta_{jk}-\delta_{ik}\delta_{mj})\partial_i w_{jk} \\
    &= \int_S dA\, (\partial_m w_{kk} - \partial_k w_{mk}). \label{eqn_deriv_2}
\end{align}
The vector on the left-hand side, obtained by rotating the Burgers 
vector by $\pi/2$ counterclockwise, is the dipole 
moment associated with the pair of disclinations that make up the 
dislocation, which we call $\mathbf{d}$. For an elementary 
dislocation in the hexagonal lattice, this is a vector of length 
$a_0$ that points from the fivefold to the sevenfold point. 
Equations~\ref{eqn_deriv_1} and \ref{eqn_deriv_2} together give the 
topological count in terms of the disclination dipole moment, 
\begin{equation}
    \nut = \frac{1}{V_\text{cell}^0} \pt^0_i \varepsilon_{ji}b_j 
         = \frac{1}{V_\text{cell}^0} \Pt \cdot \mathbf{d}.
\end{equation}

\section{Localization properties of topologically protected modes} 
\label{app_localization}
\begin{figure*}
    \includegraphics{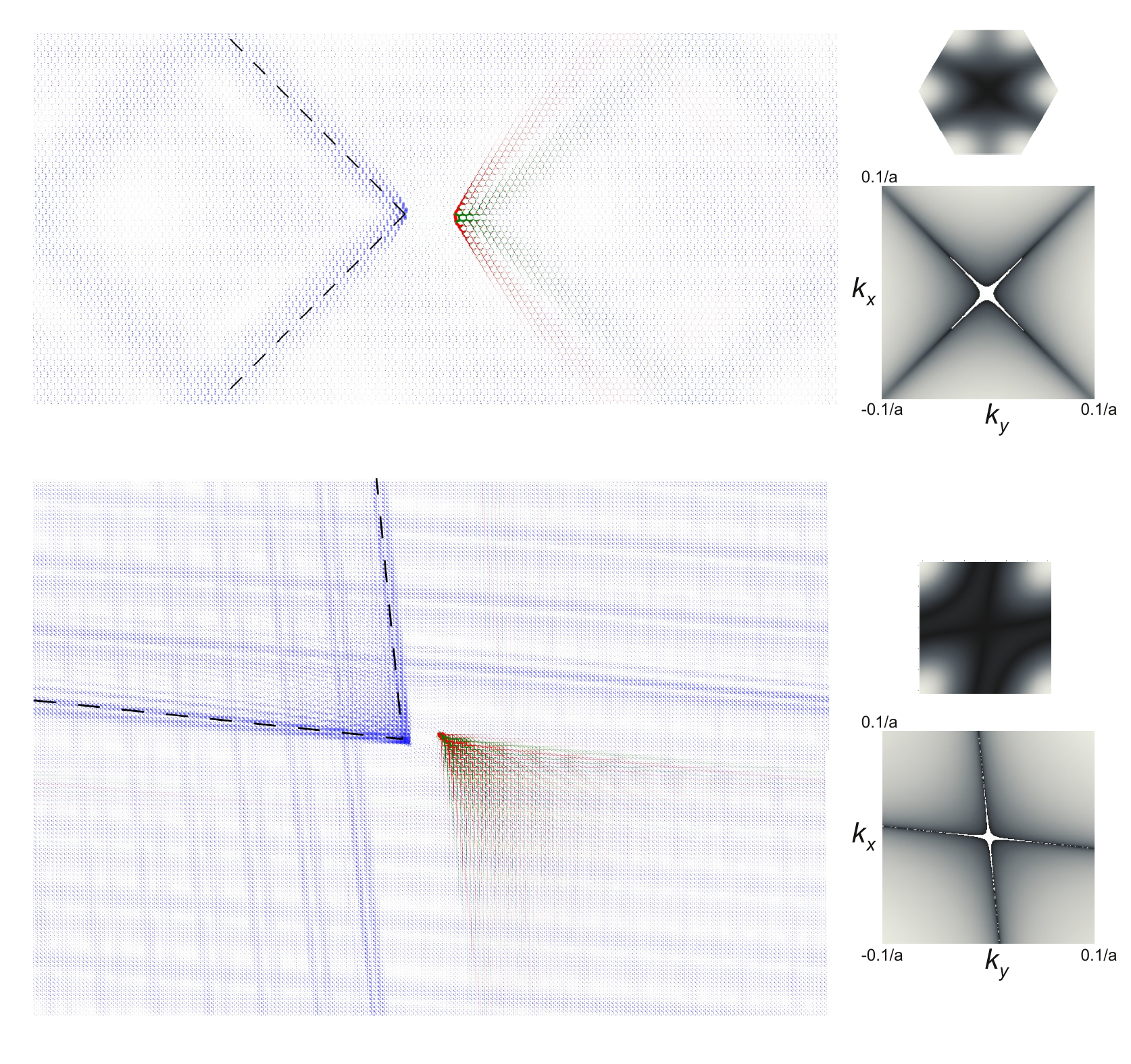}
    \caption{{\sffamily
        Zoomed-out versions of the scatter plots of the mode amplitude from 
        Fig.~3 of the main text. The entire lattice is shown for both the 
        kagome- and square-based topological lattices (consisting of 
        $128\times 64$ unit cells for the deformed kagome and $192\times 128$ 
        unit cells for the deformed square lattice). The approximate state 
        of self-stress associated with the opposite dislocation is also 
        visualized for each lattice by 
        scaling the thickness of bonds by the magnitude of the tension (green) 
        or compression (red) as in main text Fig.~2. The wrapping of the weak 
        localization direction around the periodic boundaries is apparent. To the top right of each 
        lattice is a density plot of $\det R$ as a function of $\mathbf{k}$ in 
        the respective Brillouin zone. Below this, a density plot of $\log 
        \det R$ is shown for a subsection of the Brillouin zone near the 
        origin, to highlight the directions of anomalous dispersion (sharp 
        lines) along which $\det R = 0$ to quadratic order in
$(k_x,k_y)$. The axes of the Brillouin zone have been swapped to aid
the visual comparison of the localization direction in real space and
the anomalous dispersion directions in momentum space.
    }} \label{fig_si_zmss_zoomout}
\end{figure*}

The numerical results of the main text show that, in certain spatial directions away from the
dislocation, the decay of the amplitude of the soft mode associated with a 
dislocation is much slower than in the other directions (Fig.~3, main text). 
We observe here that these directions appear to be controlled by the 
long-wavelength elastic behaviour of the bulk lattice without defects.

The authors of Ref.~\onlinecite{Kane2013} showed that in the 
long-wavelength limit of the elasticity of the deformed kagome
structures with $\Pt \neq 0$, there are two directions in the Brillouin zone which exhibit 
anomalous dispersion of the long-wavelength acoustic modes as a function of 
the wavevector $\mathbf{k}=(k_x,k_y)$. Along these 
directions, the bulk modes have
 a dispersion that scales as $\omega \sim k^2$ rather than
the typical situation for acoustic phonon modes which is $\omega \sim k$. 
 These directions are revealed by expanding the determinant of the 
 Fourier-transformed rigidity matrix $R(\mathbf{k})$ in powers of $(k_x,k_y)$; along the special 
directions, characterized by $k_x = \alpha_i k_y$ ($i \in\{1,2\}$) where 
$\alpha_1$ and $\alpha_2$ are determined by the unit cell, $\det R$ vanishes to 
quadratic order in $k$~\cite{Kane2013}. 
Therefore,  $k_x = \alpha_i k_y$ specify lines 
of  modes that are gapless to quadratic order in $k$, whose energy is made finite only by the 
presence of higher-order terms. Such near-gapless lines in 
the dispersion indicate a multitude of low-energy modes of the lattice 
whose associated displacements are constant along the special directions $y = 
\alpha_i x$ in real space. 

We observe that the directions of weak localization of the topological soft modes in 
real space tracks these special directions originating in the anomalous 
dispersion in $\mathbf{k}$-space. Fig.~\ref{fig_si_zmss_zoomout} compares the 
directions of weak localization of the topological modes (dashed lines) to the 
directions of anomalous dispersion in $\mathbf{k}$-space (sharp lines in 
density plot). As mentioned earlier, the deformed kagome lattice used for the lattices and 
prototypes studied in the main text corresponds to $(x_1,x_2,x_3)=(.12,-.06,-.06)$ 
in the parametrization of Ref.~\onlinecite{Kane2013}. For such a lattice, 
$\alpha_1 = -\alpha_2$~\cite{Kane2013} (we obtain specifically $\alpha_i = \pm \sqrt{(3 
x_1)/(x_1-4x_2)} \approx \pm 1$). We numerically obtain a similar feature 
in the determinant of the Fourier-transformed rigidity matrix of the deformed 
square lattice, with the two directions $k_x = \alpha_i k_y$ along which $\det R=0$ to  
quadratic order in $k$ quantified by $\alpha_1 \approx 9.995$ and $\alpha_2 
\approx 0.105$. In both lattices, the directions of weak localization are 
consistent with the real-space directions of soft response corresponding to 
the near-gapless lines, $y = 
\alpha_i x$ (the special directions in real space and 
$\mathbf{k}$-space are related by a $\pi/2$ rotation, thus the axes of
the Brillouin zone in Fig.~\ref{fig_si_zmss_zoomout} have been swapped
to make the correspondence more apparent). 

\begin{figure*}
    \includegraphics[width=170mm]{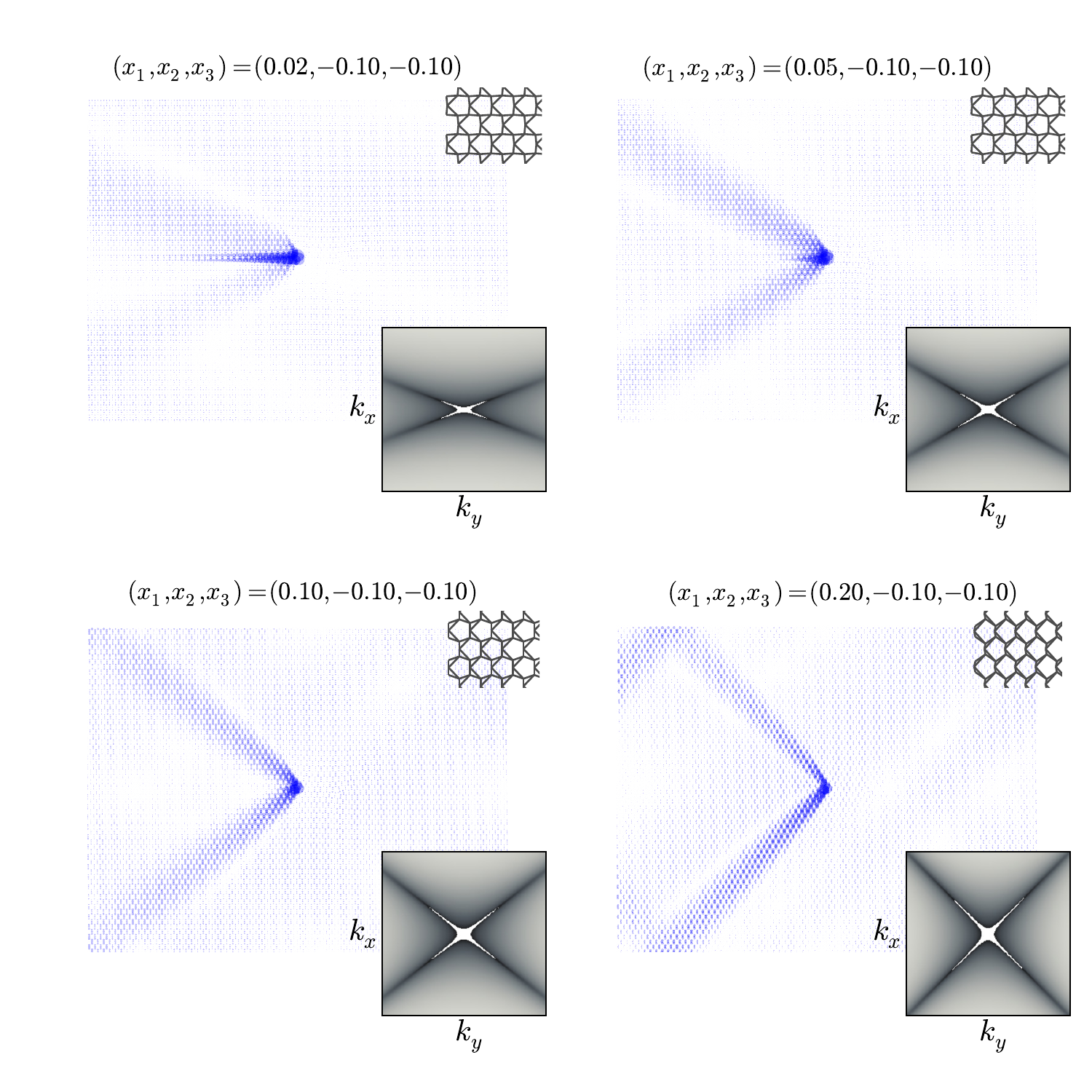}
    \caption{{\sffamily
        Scatter plots similar to Fig.~\ref{fig_si_zmss_zoomout} for the 
        numerically-obtained soft mode associated with a dislocation in four different deformed kagome 
        lattice unit cells parametrized by $(x_1,x_2,x_3)=(x_1,-0.1,-0.1)$. 
        The lattices are $128\times 64$ unit cells in size, and have a 
        dislocation and anti-dislocation with the same orientations and 
        separation as in Fig.~3a of the main text. A 
        section of the defect-free lattice is shown in the upper right corner 
        of each scatter plot to illustrate the unit cell, and a density plot of $\log\det R$ near the 
        origin of the Brillouin zone is shown in the lower right corner to 
        indicate the directions of anomalous dispersion. The density plot is 
        rendered with $k_x$ on the vertical axis to highlight the visual 
        similarity with the directions of weak localization of the soft mode 
        in real space.
    }} \label{fig_si_modeangle}
\end{figure*}

This behaviour is confirmed numerically in other members of the class of 
deformed kagome lattices of Ref.~\onlinecite{Kane2013}, parametrized by 
$(x_1,x_2,x_3)=(x_1,-0.1,-0.1)$, for which $\alpha_1 = -\alpha_2 = \sqrt{(3 x_1)/(x_1-4x_2)}$. 
Upon increasing $x_1$ from zero, the weak localization direction in real 
space tracks the change in anomalous dispersion direction in $\mathbf{k}$-space, as 
illustrated in Fig.~\ref{fig_si_modeangle}, again suggesting that the origin 
of the former lies in the latter. Establishing the theoretical relation 
between the two special directions would be an interesting avenue for future 
work.

\end{document}